\documentclass[preprint]{elsarticle}
\usepackage[utf8]{inputenc}
\usepackage{amssymb}
\usepackage{amsthm, amsmath}
\usepackage{stackengine}
\usepackage{url}
\usepackage{alltt}
\usepackage{proof}
\usepackage{xcolor}
\usepackage{eurosym}
\usepackage{xspace}

\def\eqbydef{\mathrel{\ensurestackMath{\stackon[1pt]{=}{\scriptstyle\Delta}}}}

\newtheorem{theorem}{Theorem}[section]
\newtheorem{proposition}{Proposition}[section]
\newtheorem{lemma}{Lemma}[section]
\newtheorem{example}{Example}[section]
\newtheorem{definition}{Definition}[section]
\newtheorem{notation}{Notation}[section]

\newcommand{\step}{\curvearrowright}
\newcommand{\steps}{\rightsquigarrow}
\newcommand{\tr}[6]{
     \{ #1 : #2 $\to$ #3 ; #4 {\tt #5} ; #6 \}
}
\newcommand{\ctr}[4]{
    $\langle$ #1 : #2 $\to$ #3 ; #4 $\rangle$
}

\usepackage{lineno}

\title{Certifying Findel Derivatives for Blockchain}

\author{Andrei Arusoaie}

\address{Alexandru Ioan Cuza University of Ia{\c s}i\\\texttt{andrei.arusoaie@uaic.ro}}

\begin{document}

\begin{frontmatter}



\begin{abstract}
Derivatives are a special type of financial contracts used to hedge risks or to speculate on the market fluctuations.
In order to avoid ambiguities and misinterpretations, several domain specific languages (DSLs) for specifying such derivatives have been proposed.
The recent development of the blockchain technologies enables the automatic execution of financial derivatives.
Once deployed on the blockchain, a derivative cannot be modified. 
Therefore, more caution should be taken in order to avoid undesired situations.

In this paper, we address the formal verification of financial derivatives written in a DSL for blockchain, called Findel.
We identify a list of properties that, once proved, they exclude several security vulnerabilities (e.g., immutable bugs, money losses).
We develop an infrastructure that provides means to interactively formalize and prove such  properties. To provide a higher confidence, we also generate proof certificates. 
We use our infrastructure to certify non-trivial examples that cover the most common types of derivatives (forwards/futures, swaps, options).
\end{abstract}


\end{frontmatter}


\section{Introduction}
\label{sec:intro}

Financial derivatives~\citep{steinherr1998derivatives,hull2009options,chisholm2010derivatives,gottesman2016derivatives} play an important role in finance. A financial derivative is a contract between two or more parties whose value is based on underlying financial assets (e.g., bonds, commodities, etc.).  The term \emph{derivative} captures the idea that the contract \emph{derives} its value from fluctuations in the underlying asset. For instance, a car insurance contract derives its value from an insurance index and related events (e.g.,  an accident may change its value).

Derivatives are often used to express complex agreements. Suppose that John is a cereal manufacturer and Tom is a farmer who produces wheat and corn. Since the price of cereals fluctuates a lot, John believes that it would be a very good idea to have an agreement with Tom at the beginning of the year: he computes a convenient price and makes an offer to Tom for buying his products in advance. If Tom agrees, then John will not be affected by a rise in the price at the end of the year. If the price drops, then Tom will have an advantage. Should John and Tom decide to accept this agreement, they can make accurate predictions for the future.

Typically, financial derivatives are expressed using natural language on a written document which is authorized by a trusted third party. 
However, natural language is often ambiguous and lacks precision.
This might cause disputes between the parties involved, even in the presence of a trusted authority. \begin{figure*}[htp]
\begin{center}
\fbox{\includegraphics[scale=0.6]{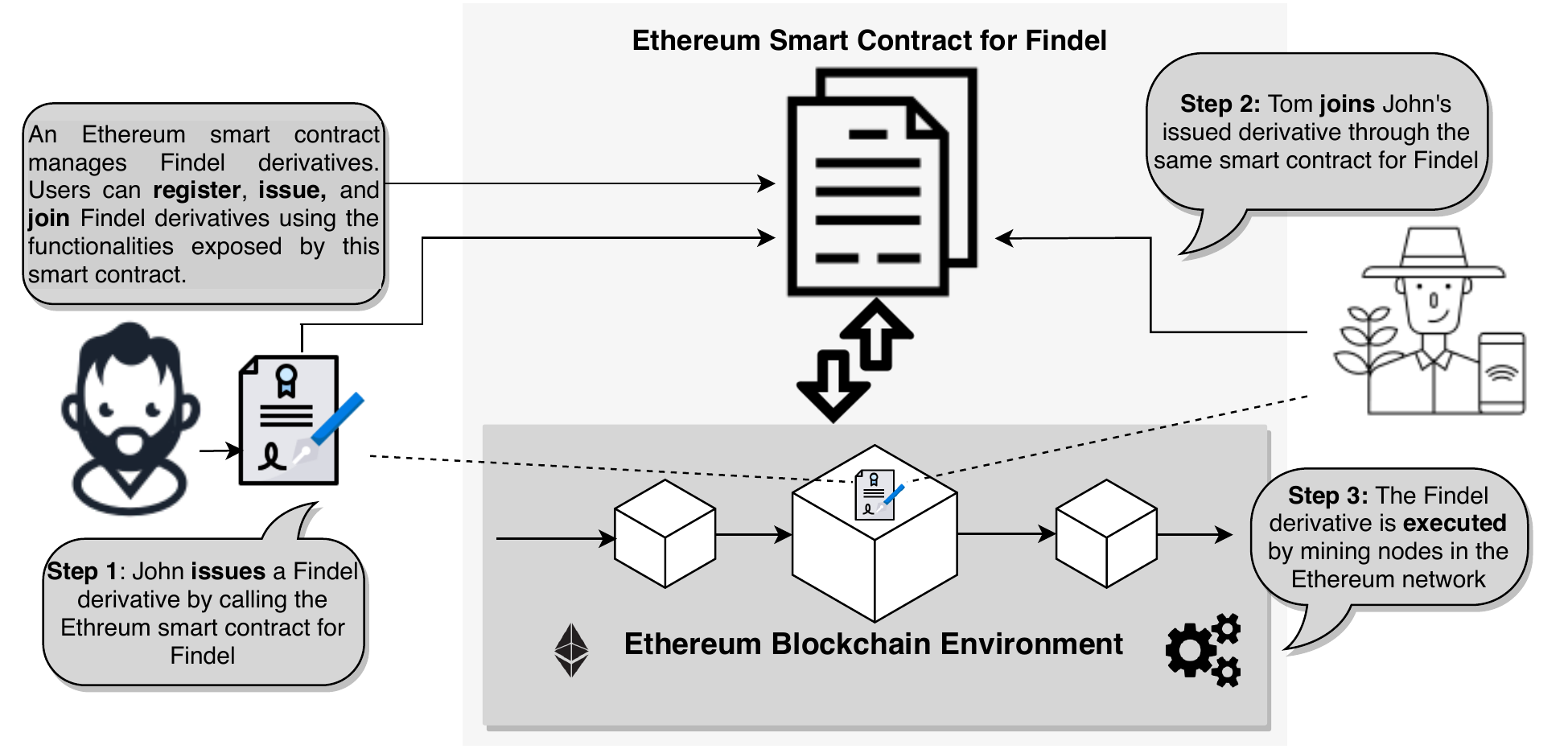}}
\end{center}
\label{fig:etherfindel}
\caption{Findel derivatives are managed by an Ethereum smart contract~\citep{findelgit} implemented in Solidity~\citep{solidity}. 
Scenario: John \emph{issues} a derivative using the Ethereum smart contract; the derivative is now available and anyone can \emph{join} it. Here, Tom \emph{joins} the derivative issued by John through the same smart contract. Tom becomes the \emph{owner} of the derivative and the derivative is executed by the miners in the network.}
\end{figure*}
\paragraph{Domain Specific Languages for financial derivatives}
In order to avoid disputes, researchers have proposed simple, non-ambiguous, and \emph{executable} domain specific languages (DSLs) for describing financial derivatives. 
\citet{jones2000composing} proposed such a language which was implemented as a combinator library in Haskell. They have defined a minimal and expressive set of \emph{primitives}, which are then used to create complex derivatives. An example is {\tt One(EUR)} - a primitive for transferring one unit of currency ({\tt EUR}) from one party to another. New derivatives can be created from basic ones: e.g., {\tt Scale(k, One(EUR))} multiplies by {\tt k} the amount to be transferred.
Derivatives expressed using primitives are \emph{composable}. For example, {\tt And($c_1$,$c_2$)} processes the derivatives $c_1$ and $c_2$ sequentially.
\citet{jones2000composing} were able to compute the value of a derivative by giving a \emph{valuation semantics} to their Haskell combinators. 




Another language in the same family has been proposed by~\citet{gai}.
It uses similar primitives as in ~\citet{jones2000composing} and has a denotational semantics. 
One of the design goals was to enable standard mathematical analysis, i.e., study whether sequences of payments are consistent with the contract descriptions. 


The blockchain technology enables \emph{automatic} processing of financial derivatives via \emph{smart contracts}. In this context, new DSLs for derivatives have emerged: Findel~\citep{findel}, Marlowe~\citep{marlowe}, and a contract language introduced by~\citet{egelund2017}.

\paragraph{Blockchain}
The blockchain technology is used mainly by cryptocurrencies such as Bitcoin~\citep{bitcoin}. It involves a peer-to-peer network of nodes that communicate with the purpose of contributing to a distributed ledger. This ledger is a chain  of blocks of transactions. Only a special category of nodes, called \emph{miners}, contribute to the creation of blocks. Their role is to arrange transactions in blocks in a way that requires computing power. Miners receive a reward as incentive. 
A block is added to the main ledger by a consensus algorithm. The consensus algorithm ensures that it is not convenient to change the existing ledger, but only to add new blocks to it. This is why the blockchain is said to be \emph{immutable}.
Various blockchain platforms such as Ethereum~\citep{wood,eth} or Cardano~\citep{cardano} allow users to create smart contracts, i.e., programs that typically handle transactions involving cryptocurrency, but they can also encode complex logic like regular programs. Miners are incentivised by an execution fee. 


\paragraph{Vulnerabilities in Findel}
In this paper we focus on Findel, a DSL based on a set of primitives which is very similar with the set proposed by~\citet{jones2000composing} and~\citet{gai}. The main difference is that Findel derivatives can be automatically executed in Ethereum (Figure~\ref{fig:etherfindel}). Findel can be regarded as a smart contracts language for financial derivatives. Therefore, Findel inherits several vulnerabilities of smart contracts.
A comprehensive taxonomy of vulnerabilities of smart contracts can be found in~\citep{Atzei:2017:SAE:3080353.3080363}. We recall here the vulnerabilties that are relevant for  Findel:
\begin{itemize}
  \setlength\itemsep{0em}
\item \emph{Immutable bugs}. Due to the immutability of the blockchain, deployed Findel code cannot be modified. This is consistent with the principles of Ethereum, where \emph{code is law}. If the Findel code contains a bug, there is no direct way to fix it. 


\item \emph{Money lost in transfer}. When sending money, one has to specify the correct recipient address. Money sent to wrong addresses are lost forever.

\item \emph{Time constraints}. In Findel, time constraints are part of the language: a contract issuer can specify time intervals when contracts can be joined. If an interested owner fails to join in time, then the owner may lose money or other contracts.
\end{itemize}

Programming smart contracts is a tricky task and developers need tools to verify whether their programs are more secure. In this context, formal verification of smart contracts has started to be a very active field \citep{survey,wang,Bhargavan:2016:FVS:2993600.2993611,kalra2018zeus,Tsankov:2018:SPS:3243734.3243780}.
Most of the tools based on formal verification focus on verifying general security properties of the Ethereum bytecode (e.g., ~\citep{10.1007/978-3-319-89722-6_10}). 

\paragraph{Handling vulnerabilities in Findel}
Many vulnerabilities are caused by a misalignment between the intuition of the programmer and the semantics of the language.
The main sources of vulnerabilities in Findel are two statements that (1) allow participants to swap the parties (issuer and owner) and (2) generate new contracts. We show here a list of properties that, once verified, could exclude certain vulnerabilities (e.g., immutable bugs, money losses) for the derivative in question:

\newcommand{\cm}{\textit{(CM)}\xspace}
\newcommand{\es}{\textit{(ES)}\xspace}
\newcommand{\as}{\textit{(AS)}\xspace}

\begin{itemize}
  \item[\cm] Derivatives should be free of \emph{calculation mistakes}. This is a typical bug in financial contracts where mistakes in mathematical formulas could lead to wrong amounts of money to be transferred. In the real world, this is can be corrected. Unfortunately, a bug in a deployed Findel derivative cannot be fixed at all, due to immutability of the blockchain.
  \item[\es] Errors caused by \emph{external sources} should be handled properly. Findel allows users to retrieve data from external sources using gateways. A simple example is a Findel currency exchange contract, where the exchange rate is provided by an external source. Suppose that Charlie wants to exchange \$10 into euros. So, he joins the issued contract and pays the \$10. If the external source fails to provide an exchange rate and the Findel code does not handle such situations properly, then Charlie receives nothing in exchange, and losses the \$10.
   \item[\as] \emph{Accidental swaps} should be avoided. This property ensures that the generated transactions and contracts have the intended issuer and the intended owner. Although this may seem easy to check, sometimes this is quite difficult to detect when statements (1) and (2) are combined. In Findel, the execution of a contract is triggered when someone \emph{joins} an issued contract. Moreover, you cannot force someone to join.
Suppose that John issues $c_1$ which specifies that: (a) he pays in advance for cereals, and (b) a new contract $c_2$ with John as issuer is generated (by $c_1$), where $c_2$ specifies that John must receive cereals in exchange. If Tom joins $c_1$, then Tom receives the payment and he is expected to join $c_2$. However, if Tom is dishonest and he does not join $c_2$, then John never gets the cereals! The problem is that the execution of $c_2$ depends on Tom, and he might not be interested in joining it.\footnote{In Section~\ref{sec:mot} we discuss this particular issue in detail.}.
  \end{itemize}

In this paper we tackle the formal verification of this kind of properties for Findel contracts. We develop an infrastructure implemented in Coq~\citep{coq}\footnote{The source code is available online: \url{https://github.com/andreiarusoaie/findel-semantics-coq}}, that provides means to execute Findel derivatives, and to formalize, prove, and certify properties about them.

Compared to the other formal verification approaches, these are mostly focused on formalizing and proving generic security properties of Ethereum bytecode ~\citep{wang, Tsankov:2018:SPS:3243734.3243780}, and only a few actually certify these properties~\citep{10.1007/978-3-319-89722-6_10, Bhargavan:2016:FVS:2993600.2993611}.

Compared to the other DSLs~\citep{jones2000composing,gai} in the same family as Findel, none of them are designed to be executed on the blockchain. 
They are mostly focused on computing derivatives values or performing conformance analyses.
Here we focus is different: we prove properties that make sense when such derivatives are executed in the blockchain.

\paragraph{Contributions}
The contributions of this paper are:
\begin{enumerate}
\item A formal semantics of Findel in Coq~\citep{coq}.
\item An infrastructure that can be used to certify security properties for the most common types of derivatives (futures/forwards, swaps, options). 
\item A list of examples that highlight the various security properties that derivatives have. We use our infrastructure to prove these properties. When a proof of a security property cannot be completed, we show how to detect the source of the problem.
\item A practical method for developing proofs for complex derivatives.
Since Findel is composable, proofs can be divided into smaller manageable pieces. 
We show that proofs of complex derivatives can be done by composing smaller proofs.
\end{enumerate}

\paragraph{Paper organisation}
Section~\ref{sec:findel} introduces the syntax and the informal semantics of Findel from~\citep{findel}. 
Section~\ref{sec:mot} contains a motivating example and we discuss what properties are violated by this example.
The Coq encoding of Findel is shown in Section~\ref{sec:coq}. 
Using the Coq semantics we encode several Findel derivatives and then we prove their expected properties  in Section~\ref{sec:experiments}. 
We also include a credit default swap - a very popular type of derivative - and we show how its proofs can be developed incrementally in Section~\ref{sec:cds}.
We conclude in Section~\ref{sec:conclusion}.

\section{Findel: a DSL for financial derivatives}
\label{sec:findel}

The syntax and the informal semantics of the Findel language are both described in~\citet{findel}. Details can be found in the implementation available at \url{https://github.com/cryptolu/findel}. In this section we use both sources in order to make a complete description of the language. We do not improve Findel here, we only to formalise its existing semantics.

A \emph{Findel contract} is a tuple with three components: a \emph{description}, an \emph{issuer} and an \emph{owner}. The \emph{issuer} and the \emph{owner} are the parties of the given contract. In the implementation, the issuer and the owner are represented as 20-byte values (i.e., size of an Ethereum account address), and contracts have an additional component called \emph{proposed owner} which is also a 20-byte value. This new component is used to propose a specific owner when issuing a new contract. If the proposed owner field is {\tt 0x0} (default value), then anyone can join the contract. 

A Findel \emph{description} is essentially a tree with \emph{basic primitives} as leaves and \emph{composite primitives} as internal nodes. The list of available primitives and their informal semantics is shown in Table~\ref{tbl:informal}.

\begin{table}[tp]
\small
\centering
\begin{tabular}{| l | l | }
\hline
\bf Primitive & \bf Informal semantics\\
\hline\hline
{\tt Zero} & Do nothing.\\
\hline

{\tt One(}\textit{currency}{\tt )} & Transfer 1 unit of {\textit{currency}} from \\
& the issuer to the owner.\\
\hline
\hline
{\tt Scale(}$k$, $c${\tt )} & Multiply all payments of $c$ by a \\ 
& constant factor of $k$.\\
\hline
{\tt ScaleObs(}$a$, $c${\tt )} & Multiply all payments of $c$ by a \\ 
& factor obtained from address $a$.\\
\hline
{\tt Give(}$c${\tt )} & Swap parties of $c$.\\
\hline
{\tt And(}$c_1$, $c_2${\tt )} & Execute $c_1$ and then $c_2$.\\
\hline
{\tt Or(}$c_1$, $c_2${\tt )} & Give the owner the right to execute\\
& either $c_1$ or $c_2$ but not both\\
\hline
{\tt If(}$a$, $c_1$, $c_2${\tt )} & If $b$ is true, execute $c_1$, else execute\\
& $c_2$, with $b$ obtained from address $a$.\\
\hline
{\tt Timebound(}$t_0$, $t_1$, $c${\tt )} & Execute $c$, if the current timestamp \\
&  is within $[t_0, t_1]$.\\
\hline
\end{tabular}
\caption{The informal semantics of Findel~\citep{findel}.}
\label{tbl:informal}
\end{table}


\begin{example}
\label{ex:frce}
{\tt And} is a primitive that executes two (sub)contracts, $c_1$ and $c_2$, sequentially. If at least one of them fails then the changes are reverted.
Here is a fixed-rate currency exchange derivative in Findel:
\begin{center}
{\tt
\small
And(Give(Scale(11, One(USD))), \\Scale(10, One(EUR)))}.
\end{center}
\end{example}

To increase the expressivity of Findel, the language is extended with some sugar syntax as shown in Table~\ref{tbl:additional}.
\noindent
\begin{table}[t]
\centering
\small
\begin{tabular}{|l|l|}
\hline
    {\bf Additional syntax} & {\bf Desugared syntax} \\
    \hline\hline
    {\tt At($t_0$, $c$)} & {\tt Timebound($t_0 - \delta$, $t_0 + \delta$, $c$}),\\
    & where $\delta$ is just a constant used\\
    & to handle the imperfect precision\\
    & of time signals in the network\\
    \hline
    {\tt Before($t_0$, $c$)} & {\tt Timebound($now$, $t_0$, $c$)}, \\ 
    & where $now$ is the current time \\
    \hline
    {\tt After($t_0$, $c$)} & {\tt Timebound($t_0$, $\infty$, $c$)}. \\
\hline
\end{tabular}{}
\label{tbl:additional}
\caption{Additional primitives}
\end{table}

\begin{example}
\label{ex:zcb}
A zero-coupon bond (\textit{ZCB}) can be encoded using {\tt At}:\\[1ex]
\noindent
{\tt \small \hspace*{1cm}And(Give(Scale(10, One(USD))),\\ \hspace*{2cm}At(\textit{now}+t, Scale(11, One(USD))))}
\end{example}

The \emph{execution model} of a Findel contract is given by these steps~\citep{findel}:
\begin{enumerate}
\setlength{\itemsep}{0mm}
\item The first party \emph{issues} a contract. This is a mere declaration of the issuer's desire to conclude an agreement and entails no obligations.
\item The second party \emph{joins} the contract and becomes its owner. As a consequence, both parties accept the specified rights and obligations.
\item The contract is executed immediately:
	\begin{enumerate}
		\item Let the current node be the root  of the contract description.
				If the current node is either {\tt Or} or {\tt Timebound} with $t_0 > \mathit{now}$, postpone the execution: issue a new contract, with the same parties and the current node as root. The owner can demand later its execution.
		\item Otherwise, execute all sub-nodes recursively.
		\item Delete the contract.
	\end{enumerate}
\end{enumerate}

\begin{example}
\label{ex:execution}
A step by step execution of the contract in Example~\ref{ex:frce}, where Alice issues and Bob is the owner:
\begin{enumerate}
    \item Alice issues the contract and Bob joins it, becoming the contract owner:\\
    {\small
    \begin{tabular}{|c|c|c|l|}
    
        \hline
        {\bf  } & \$ & \euro & {\bf is owner of} \\
        \hline
        Alice & 100 & 50 & - \\
        \hline
        Bob & 20 & 30 & {\tt And(Give(}\\
        &&& {\tt ~~~~~Scale(11, One(USD)))} \\
        &&& {\tt  ~~~~Scale(10, One(EUR)))}\\
        \hline
    \end{tabular}}
    
    \item {\tt And} executes and Bob is now the owner of two (sub)contracts:\\
    {\small
    \begin{tabular}{|c|c|c|l|}
        \hline
        {\bf } & \$ & \euro & {\bf is owner of} \\
        \hline
        Alice & 100 & 50 & - \\
        \hline
        Bob & 20 & 30 & {\tt Give(Scale(11, One(USD)))} \\
        &&& {\tt  Scale(10, One(EUR))}\\
        \hline
    \end{tabular}
    }

    \item {\tt Give} executes and Alice becomes an owner:\\[1ex]
    {\small
    \begin{tabular}{|c|c|c|l|}
        \hline
         & \$ & \euro & {\bf is owner of} \\
        \hline
        Alice & 100 & 50 & {\tt Scale(11, One(USD))} \\
        \hline
        Bob & 20 & 30 & {\tt  Scale(10, One(EUR))}\\
        \hline
    \end{tabular}
    }
    
        \item Alice receives 11 {\tt USD} from Bob:\\[1ex]
    {\small
    \begin{tabular}{|c|c|c|l|}
        \hline
        & \$ & \euro & {\bf is owner of} \\
        \hline
        Alice & 111 & 50 & - \\
        \hline
        Bob & 9 & 30 & {\tt Scale(10, One(EUR))} \\
        \hline
    \end{tabular}
    }

    \item Finally, Bob receives 10 {\tt EUR} from Alice:\\[-1ex]

    {\small
    \begin{tabular}{|c|c|c|l|}
        \hline
         & \$ & \euro & {\bf is owner of} \\
        \hline
        Alice & 111 & 40 & - \\
        \hline
        Bob & 9 & 40 & - \\
        \hline
    \end{tabular}
    }
    
\end{enumerate}

\end{example}

The implementation of Findel does not enforce any constraints on the balances of the users that prevents them from building up debt.

\subsection{Limitations of Findel}
Findel has several major limitations. First, there is no way to encode repetitive behavior in contracts - there are no loops. This is a limitation when one wants to specify a contract should be executed repeatedly for 10 years.

Second, by design, Findel is not able to express agreements for more than two parties. One can access external gateways to simulate multiple parties, but this is a very limited workaround.
\subsubsection{A motivating example}
\label{sec:mot}


Recall our cereal manufacturer example from Section~\ref{sec:intro}. Suppose that John issues the following option derivative (here, wheat is represented by {\tt USD} and corn by {\tt EUR}):

\noindent
\begin{minipage}{\linewidth}
\small
\tt 
\begin{tabular}{c c l}
  $\mathit{OPT} \eqbydef$&And\big(& \\
     && \hspace*{-2ex}Before(t,Or(Give(One USD),\\
     && ~~~~~~~~~~Give(One EUR)))\\
     && \hspace*{-2ex}After(t+2, Scale(1, (One GBP)))\big)
\end{tabular}{}
\end{minipage}

\noindent
John wants to acquire a contract before moment {\tt t} that gives him a later choice (between wheat and corn). He pays back 1 {\tt GBP} as incentive for eventual owners after {\small\tt t+2}.
Unfortunately, John issues a contract with bugs.

Suppose that a dishonest participant, Mallory, joins the contract and triggers the execution of {\tt And}. First, {\tt Before} is executed and its enclosed contract (i.e., {\tt Or}) produces a new contract - a future option - whose execution can be triggered later by Mallory.
Second, the {\tt After} primitive gets executed, and John will pay 1 {\tt GBP} after moment {\tt t+2}. 

The contract generated by {\tt After} has all the aforementioned properties: \cm, \es, and \as. Mallory is able to request 1 {\tt GBP}  from John after {\tt t+2}, and John cannot deny or retract from that.

Unfortunately,  the future option contract generated by {\tt Or} violates \as: because Mallory is the owner of this contract, she decides not to join that contract. So, John cannot claim anything from Mallory and losses money.

In Section~\ref{sec:experiments}, we show a possible fix of $\mathit{OPT}$ and we prove that in the fixed contract John has the option to choose the products.


\section{A formal semantics of Findel in Coq}
\label{sec:coq}


\subsection{Syntax}
\label{sec:syntax}
The syntax of Findel is fairly small in size and it is encoded in Coq using {\tt Inductive} (Figure~\ref{fig:syntax}):

\begin{figure}[htp]
\begin{center}
\begin{minipage}{.7\linewidth}
\small
\begin{alltt}
Inductive Primitive :=
(* basic primitives *)
| Zero     :                                  -> Primitive
| One      :Currency                          -> Primitive
(* composite primitives *)
| Scale    :nat -> Primitive                  -> Primitive 
| ScaleObs :Address -> Primitive              -> Primitive
| Give     :Primitive                         -> Primitive
| And      :Primitive -> Primitive            -> Primitive
| Or       :Primitive -> Primitive            -> Primitive
| If       :Address -> Primitive -> Primitive -> Primitive
| Timebound :nat->nat -> Primitive            -> Primitive.
\end{alltt}
\end{minipage}
\caption{The syntax of Findel in Coq.}
\label{fig:syntax}
\end{center}
\end{figure}

Currencies are defined using {\tt Inductive} as well:

\begin{center}
\begin{minipage}{.5\linewidth}
\small
\begin{alltt}
Inductive Currency :=
| USD  : Currency
| EUR  : Currency
| GBP  : Currency
| JPY  : Currency
| CNY  : Currency
| SGD  : Currency
| NONE : Currency.
\end{alltt}
\end{minipage}
\end{center}

\noindent
The additional primitives {\tt At}, {\tt Before}, {\tt After}, and {\tt Sell} are simple definitions:\\

\begin{minipage}{\textwidth}
\small
{\tt Definition At (t:nat) (p:Primitive)    := }{\tt Timebound (t - $\delta$) (t + $\delta$) p.}\\
{\tt Definition Before (t:nat) (p:Primitive) :=}{\tt Timebound 0 t p.}\\
{\tt Definition After (t:nat) (p:Primitive)   :=}{\tt Timebound t INF p. }\\
\end{minipage}

\noindent
$\delta$ is a just parameter which is used to adjust intervals for accepting transactions.
The infinite is axiomatised in Coq as follows:\\

\begin{minipage}{\linewidth}
{\tt \small
Parameter INF : nat.\\
Axiom infinite : forall n, n < INF.
}
\end{minipage}

\begin{example}
When written in Coq, contracts look very similar to what they looked before.
The contract shown in Section~\ref{sec:mot}, is encoded in Coq as follows:\\

\begin{minipage}{\linewidth}
\small
\tt 
\begin{tabular}{c c l}
  $\mathit{OPT} \eqbydef$&(And& \\
     && (Before t (Or (Give (One USD),\\
     && ~~~~~~~~~~~~(Give (One EUR))))\\
     && (At t+2 (Scale 1 (One GBP))))
\end{tabular}{}
\end{minipage}

%
%
%
\end{example}

\subsection{The semantics of Findel primitives}
\label{sec:semantics1}

As mentioned in~\citep{findel}, contracts are executed recursively. In~\citep{findelgit}, a function called {\tt execute} calls an internal function {\tt executeRecursive} that actually executes the contract recursively. The {\tt Or} primitive needs an extra-argument provided by the owner, and this is handled by yet another function {\tt executeOr} which calls {\tt execute} on the primitives specified by the extra-argument.


\paragraph{Addresses. Time. Balance} 
An address in Findel is represented by a 20-byte value. From our perspective, these are just numbers, and this is why we model addresses as naturals. We make the following convention: 0 is treated as the constant {\tt 0x0}, i.e, 0 is the default address.
The time is represented as the number of seconds that passed since a given timestamp. The balance is a function which takes an address and a currency, and returns an integer value representing the amount of tokens of the specified currency. 
An update function for balance is also provided: it takes an existing balance, an address, a currency and an amount, and produces a new balance.


\paragraph{Gateway}

\emph{Gateways} are the solution that the designers of Findel found to model interactions with external data providers. Gateways are smart contracts that provide a value, a timestamp, and a proof of authenticity. 
Before executing a contract, the gateway should be always updated~\citep{findel}.
In Coq, a gateway is a triple holding an address, a value and a timestamp:\\[-1ex]

\begin{minipage}{\linewidth}
\small
\begin{alltt}
Record Gateway :=
  gateway \{
      gtw_addr : Address;
      gtw_value : nat;
      gtw_timestamp : nat
    \}.
\end{alltt}
\end{minipage}

~\\
\noindent
Contracts have access to a list of such triples. When a contract queries an address, the address is searched in the list, and the corresponding value and timestamp are returned. If the address is not found, then the execution of the contract fails. This functionality is provided by a Coq function called {\tt query}. 
This function also checks for the freshness of the given data, and fails if the difference between the current time and the provided timestamp is less than a threshold\footnote{In~\citep{findelgit} this threshold is 30 seconds.}. The proof of authenticity mechanism is not yet implemented in~\citet{findelgit} and we do not handle it in Coq either.

\paragraph{Transactions}
For every transfer of tokens a transaction is registered in a ledger. A transaction is a uniquely identified tuple containing: the id of the contract which generated the transfer, the addresses of the participants, the amount of currency transferred, and a timestamp:\\[-1ex]

\begin{minipage}{\linewidth}
\small
\begin{alltt}
Definition Id := nat.
Record Transaction :=
  transaction \{
      tr_id : Id;
      tr_ctr_id : nat;
      tr_from: Address;
      tr_to : Address;
      tr_amount : nat;
      tr_currency : Currency;
      tr_timestamp : Time 
    \}.
\end{alltt}
\end{minipage}

\begin{notation}
We use \tr{tx}{I}{O}{V}{C}{c} to denote a transaction $\mathit{tx}$ where $(\mathtt{tr\_from}~\mathit{tx})=I$, $(\mathtt{tr\_to}~\mathit{tx})=O$, $(\mathtt{tr\_amount}~\mathit{tx})=V$, $(\mathtt{tr\_currency}~\mathit{tx})=\mathtt{C}$, and $(\mathtt{tr\_ctr\_id}~\mathit{tx})=\mathit{id}(c)$, where $c$ generated $\mathit{tx}$.
\end{notation}

\paragraph{Contract descriptions vs. Findel contracts}
Contract descriptions are separate from contract instances. A description only defines \emph{what} a contract is, not \emph{how} it is executed~\citep{findel}. In Coq, a contract description has a unique id, and contains the code (a tree of primitives), a scale factor, a gateway, and a time interval specifying when this description is valid:\\[-1ex]

\begin{minipage}{\linewidth}
\small
\begin{alltt}
Record ContractDescription :=
  description \{
      dsc_id : Id;
      dsc_prim : Primitive;
      dsc_scale : nat;
      dsc_gateway_of : list Gateway;
      dsc_valid_from : Time;
      dsc_valid_until : Time;
    \}.
\end{alltt}
\end{minipage}

\begin{notation}
Let us consider that $d$ is a contract description. In the rest of the section we use the following notations $\mathit{id}(d) \eqbydef (\mathtt{\small dsc\_id}~d), \mathit{prim}(d) \eqbydef (\mathtt{\small dsc\_prim}~d), \mathcal{G}(d) \eqbydef (\mathtt{\small dsc\_gateway\_of}~d), \mathit{sc}(d) \eqbydef (\mathtt{\small dsc\_scale}~d)$, $\mathit{vfrom}(d) = (\mathtt{\small dsc\_valid\_from} d)$, $\mathit{vuntil}(d) = (\mathtt{\small dsc\_valid\_until} d)$.
\end{notation}

According to~\citep{findel}, Findel contracts are tuples consisting of a description and two addresses (the issuer and the owner). However, for execution, we need more ingredients, as indicated by the implementation in~\citep{findelgit}:\\ 

\begin{minipage}{\linewidth}
\small
\begin{alltt}
Record FinContract :=
  finctr \{
      ctr_id : Id;
      ctr_desc_id : Id;
      ctr_primitive : Primitive;
      ctr_issuer : Address;
      ctr_owner : Address;
      ctr_proposed_owner : Address;
      ctr_scale : nat;
    \}.
\end{alltt}
\end{minipage}
~\\
\noindent
When a Findel contract is issued, the description id, the code and the scale are initialised with the corresponding fields from the description. The proposed owner is either the default value or it is set to a particular address. 

\begin{notation}
Let $c$ be a contract. We use the following notations $\mathit{id}(c) \eqbydef (\mathtt{\small ctr\_id}~c)$, $\mathit{dsc}(c) \eqbydef (\mathtt{\small ctr\_desc\_id}~c)$, $\mathit{prim}(c) \eqbydef (\mathtt{\small ctr\_primitive}~c)$, $\mathit{issuer}(c) \eqbydef (\mathtt{\small ctr\_issuer}~c)$, $\mathit{owner}(c) \eqbydef (\mathtt{\small ctr\_owner}~c)$, $\mathit{po}(c) \eqbydef (\mathtt{\small ctr\_proposed\_owner}~c)$, $\mathit{sc}(c) \eqbydef (\mathtt{\small ctr\_scale}~c)$.
\end{notation}

\begin{notation}
We use \ctr{c}{I}{O}{P} to denote a contract $c$ where $\mathit{issuer}(c)=I$, $\mathit{po}(c)=O$, and $\mathit{prim}(c)=P$.
\end{notation}

\paragraph{Results} A \emph{result} stores the outcome of the execution of a contract. It contains the updated balance, the issued contracts, the next available id, and the updated ledger:\\[-1ex]

\begin{minipage}{\linewidth}
\small
\begin{alltt}
Record Result :=
  result \{
      res_balance : Balance;
      res_contracts : list FinContract;
      res_next : Id;
      res_ledger : list Transaction
    \}.
\end{alltt}
\end{minipage}

\subsubsection{The {\tt execute} function.}
\label{par:execute}
The code of a Findel contract is executed recursively. In Coq we define a function called {\tt execute} which corresponds to the {\tt executeRecursive} function from~\citep{findelgit}. The  inputs of {\tt execute} are:
\begin{itemize} 
\setlength{\itemsep}{0ex}
    \item the primitive to be executed;
    \item the scale factor;
    \item the addresses of the issuer and the owner;
    \item the balance of the participants;
    \item the current time (received from the network);
    \item the list of available gateways;
    \item the contract id and the description id;
    \item the next available fresh identifier - which is used to assign identifiers to the generated contracts, if any;
    \item the ledger, i.e., a list of transactions.
\end{itemize}

%

The output of {\tt execute} is of type {\tt Result}. The execution of a contract can produce a new balance, can generate new contracts, can compute new fresh ids, and can register new transactions in the ledger. The fresh id generation mechanism is quite simple: {\tt execute} takes a fresh id as input and uses it as a seed for generating unique identifiers for new contracts. In Section~\ref{sec:metaprops} we prove that our id generation mechanism is consistent, i.e., the generated ids are unique.


The {\tt execute} function is recursive on {\tt P}:
\begin{alltt}
match P with
\end{alltt}

\begin{itemize}
\setlength{\itemsep}{0mm}
    \item if {\tt P} is {\tt Zero}, then nothing changes:
    
    \item[]
    \begin{minipage}{\textwidth}
    \small
        \begin{alltt}
| Zero => Some (result balance [] n ledger)
        \end{alltt}
    \end{minipage}

    \item if {\tt P} is {\tt One}, the balance of the participants is updated, new transactions are added to the ledger, a new fresh id is generated:
    
    \item[]
    \begin{minipage}{\textwidth}
    \small
        \begin{alltt}
| One currency => Some 
   (result
     (update 
       (update balance I currncy 
         ((balance I crncy)-(Z_of_nat scale))
       )
       O currency 
       ((balance O crncy)+(Z_of_nat scale))
     ) [] (S nextId)
     ((transaction nextId ctr_id I O scale
                   crncy time) :: ledger)
    )
         \end{alltt}
    \end{minipage}
    \item if {\tt P} is {\tt Scale}, then scale the value of the  contract:
    
    \item[]
    \begin{minipage}{\textwidth}
    \small
        \begin{alltt}
| Scale k c =>
    (execute c (scale * k) I O balance time 
             gtw ctr_id dsc_id n ledger)
        \end{alltt}
    \end{minipage}

    \item if {\tt P} is {\tt ScaleObs}, then the scale for the contract is updated only if the gateway query does not fail:
    
    \item[]
    \begin{minipage}{\textwidth}
    \small
        \begin{alltt}
| ScaleObs addr c =>
    match (query gtw addr time) with
    | None => None
    | Some k =>
      (execute c (scale * k) I O balance time 
               gtw ctr_id dsc_id n ledger)
    end
        \end{alltt}
    \end{minipage}
    
    \item if {\tt P} is {\tt Give}, then swap the issuer and the owner:
    
    \item[]
    \begin{minipage}{\textwidth}
    \small
        \begin{alltt}
| Give c =>
    (execute c scale O I balance time 
             gtw ctr_id dsc_id n ledger)
        \end{alltt}
    \end{minipage}
    
    \item if {\tt P} is {\tt And}, then execute the contracts sequentially; {\tt And} fails if at least one of its subcontracts fails:
    
    \item[]
    \begin{minipage}{\textwidth}
    \small
        \begin{alltt}
| And c1 c2 =>
   match (execute c1 scale I O balance time 
            gtw ctr_id dsc_id n ledger) 
   with
   | None => None
   | Some (result bal1 Is1 n1 ledger1) =>
     match (execute c2 scale I O bal1 time 
              gtw ctr_id dsc_id n1 ledger1) 
     with
     | None => None
     | Some (result bal2 Is2 n2 ledger2) =>
       Some 
       (result bal2 (Is1 ++ Is2) n2 ledger2)
     end
   end
        \end{alltt}
    \end{minipage}

    \item if {\tt P} is {\tt If}, then the execution is determined by the gateway: if the gateway query fails, then the execution fails; if the value is 0 then execute the second contract; otherwise, execute the first contract.
    
    \item[]
    \begin{minipage}{\textwidth}
    \small
        \begin{alltt}
| If addr c1 c2 =>
   match (query gtw addr time) with
   | None => None
   | Some v =>
     if beq_nat v 0
     then (execute c2 scale I O balance time 
             gtw ctr_id dsc_id n ledger)
     else (execute c1 scale I O balance time 
             gtw ctr_id dsc_id n ledger)
   end
        \end{alltt}
    \end{minipage}

    \item if {\tt P} is {\tt Timebound}, then execute the contract only if the current time is in the time interval; if the current time is less than the inferior limit of the interval, then a new contract is issued and added to the list of generated contracts; otherwise, the execution fails:
    
    \item[]
    \begin{minipage}{\textwidth}
    \small
        \begin{alltt}
| Timebound t0 t1 p =>
  if (t1 <? time)
  then None
  else
   if (t0 <? time)
   then (execute p scale I O balance time 
           gtw ctr_id dsc_id n ledger)
   else Some 
   (result balance
    [(finctr (S n) dsc_id (Timebound t0 t1 p) 
             I O O scale)] 
    (S (S n)) ledger)

        \end{alltt}
    \end{minipage}

        \item if {\tt P} is {\tt Or}, then a new contract is issued with {\tt Or} as root; the owner can later demands its execution.
    
    \item[]
    \begin{minipage}{\textwidth}
    \small
        \begin{alltt}
| Or c1 c2 => Some 
  (result balance
          [(finctr (S n) dsc_id (Or c1 c2) 
                   I O O scale)] 
          (S (S n)) ledger)
        \end{alltt}
    \end{minipage}
\end{itemize}

\subsection{Execution model}
The execution of a Findel contract may generate new contracts. We need to model a ledger of transactions, the balance of the users, an other components needed to run contracts.
We define a general system state as a tuple $\langle \mathcal{C}, \mathcal{D}, \mathcal{B}, t, \mathcal{G}, i, \mathcal{L}, \mathtt{E} \rangle$, which holds: the list of issued contracts $\mathcal{C}$, the list of available contract descriptions $\mathcal{D}$, the balance for each user $\mathcal{B}$, the current global time $t$, a list of available gateways $\mathcal{G}$, the next fresh identifier $i$, a ledger $\mathcal{L}$ and a list of events $\mathtt{E}$. In Coq, a state is represented as follows:\\

    \begin{minipage}{\textwidth}
    \small
        \begin{alltt}
Record State :=
  state \{
      m_contracts : list FinContract;
      m_descriptions : ContractDescriptions;
      m_balance : Balance;
      m_global_time : Time;
      m_gateway : list Gateway;
      m_fresh_id : Id;
      m_ledger : list Transaction;
      m_events : list Event
    \}.
        \end{alltt}
    \end{minipage}

\begin{notation}
We use the following notations (where $s$ a {\tt State}. ): $\mathcal{C}(s) \eqbydef (\mathtt{\small m\_contracts}~s),$ $\mathcal{D}(s) \eqbydef (\mathtt{\small m\_descriptions}~s),$ $\mathcal{B}(s) \eqbydef (\mathtt{\small m\_balance}~s),$ $\mathtt{E}(s) \eqbydef (\mathtt{\small m\_events}~s), \mathcal{L}(s) \eqbydef (\mathtt{\small m\_ledger}~s),$ $\mathcal{G}(s) \eqbydef (\mathtt{\small m\_gateway}~s),$ $\mathit{fresh}(s) = (\mathtt{m\_fresh\_id}~s),$ $ \mathit{time}(s) = (\mathtt{m\_global\_time}(s))$.
\end{notation}

The {\tt State} is essentially holding all the ingredients of an online marketplace for Findel derivatives. The marketplace evolves as specified by the following rules:\\

\noindent
{\bf [Issue]} When issued, a contract $c$ is added to the list of issued contracts having a unique id $i$. Initially, the owner field contains the address of the issuer, while the proposed owner field contains the address of the intended owner. If the value of the proposed owner field is 0, then anyone can join this contract.
$\mathit{prim}(c)$ is initialised from an existing description. Also an event {\tt IssuedFor} is triggered, and the global fresh id is incremented:
$$
\infer[\mathit{d} \in \mathcal{D}]{
\langle  c : \mathcal{C}, \mathcal{D}, \mathcal{B}, t, \mathcal{G}, i + 1, \mathcal{L}, (\mathtt{\small IssuedFor}~O~i) : \mathtt{E} \rangle
}{
\langle \mathcal{C}, \mathcal{D}, \mathcal{B}, t, \mathcal{G}, i, \mathcal{L}, \mathtt{E} \rangle
},
$$

\noindent
where $c$ is a contract with $\mathit{id}(c) = i$, $\mathit{dsc}(c) = \mathit{id}(d)$, $\mathit{prim}(c) = \mathit{prim}(d)$, $\mathit{issuer}(c) = I$, $\mathit{owner}(c) = I$, $\mathit{po}(c) = O$, $\mathit{sc}(c) = \mathit{sc}(d)$.
The rule essentially says that the state above the line changes into the state below the line. Also, there is a side condition $d \in \mathcal{D}$ which needs to be fulfilled in order for the rule to apply.
We used `$:$' to denote the cons list constructor, i.e., $c$ is added in front of the list of contracts $\mathcal{C}$. \\

\noindent
{\bf [Join]} Joining a contract $c$ is the most complex operation and requires several conditions. First, $(A)$ the owner $O$ who wants to join is either the proposed owner, that is, $\mathit{po}(c) = O$, or anyone can join, that is, $\mathit{po}(c) = 0$. Second, $(B)$ the root node of the contract primitive should not be an {\tt Or}:  $\mathit{prim}(\mathit{c})  \neq \mathtt{(Or~\_~\_)}$. Third, $(C)$ the execution of $c$ is limited within a time interval given by its corresponding description $d$: $\mathit{vfrom}(d) \leq t \leq \mathit{vuntil}(d)$. Here, $t$ is the current time.
Finally, $(D)$ the execution of the associated primitive should be successful, that is, 
$\mathtt{execute}$($\mathit{prim}(c)$, $\mathit{id}(c)$, $\mathit{dsc}(c)$, $\mathit{c}(c)$, $\mathit{issuer}(c)$, $\mathit{O}$, $\mathit{t}$, $\mathcal{G}$, $\mathcal{L}$, $i$) = $\langle \mathcal{B'}, \mathcal{C'}, i', \mathcal{L'} \rangle$.
The tuple $\langle\mathcal{B'}, \mathcal{C'}, i', \mathcal{L'} \rangle$ is a {\tt Result}, where $\mathcal{B'}$ is the updated balance, $\mathcal{C'}$ is the list of contracts generated by the primitive, $i'$ is the new fresh identifier, and $\mathcal{L'}$ is the updated ledger.
The rule for joining a contract is shown below; note that {\tt E} is enriched with event $e = \mathtt{Executed}~\mathit{id}(c)$:
$$
\infer[(A), (B), (C), (D)]{
\langle \mathcal{C} \setminus \{c\} \cup \mathcal{C'}, \mathcal{D}, \mathcal{B'}, t, \mathcal{G'}, i', \mathcal{L'}, e : \mathtt{E} \rangle
}{
\langle c \in \mathcal{C}, \mathcal{D}, \mathcal{B}, t, \mathcal{G}, i, \mathcal{L}, \mathtt{E} \rangle 
}
$$

\noindent
Note that we use $c \in \mathcal{C}$ to denote that $c$ is in the list of contracts, and $\mathcal{C} \setminus \{ c\}$ to denote that $c$ is removed from $\mathcal{C}$.\\

\noindent
{\bf [Join {\tt OR}]} This rule handles the case when $(B')$ $\mathit{prim}(c)  = \mathtt{(Or}~c_1~c_2\mathtt{)}$.
In this case, the owner can execute either $c_1$ or $c_2$.
Let $\square$ be a placeholder variable which can be either $c_1$ or $c_2$.
If the execution is successful, that is, $(D')$
$\mathtt{execute}$($\square$, $\mathit{id}(c)$, $\mathit{dsc}(c)$, $\mathit{c}(c)$, $\mathit{issuer}(c)$, $\mathit{O}$, $\mathit{t}$, $\mathcal{G}$, $\mathcal{L}$, $i$) = $\langle \mathcal{B'}, \mathcal{C'}, i', \mathcal{L'} \rangle$, 
then the rule below applies:
$$
\infer[(A), (B'), (C), (D')]{
\langle \mathcal{C} \setminus \{ c\}\cup \mathcal{C'}, \mathcal{D}, \mathcal{B'}, t, \mathcal{G'}, i', \mathcal{L'}, e : \mathtt{E} \rangle
}{
\langle c\in \mathcal{C}, \mathcal{D}, \mathcal{B}, t, \mathcal{G}, i, \mathcal{L}, \mathtt{E} \rangle 
}
$$
Note that the conditions $(A)$ and $(C)$ need to be satisfied here as well. Again, $e = \mathtt{Executed}~\mathit{id}(c)$.\\

\noindent
{\bf [Fail]} If the execution of a contract fails, that is, $(D'')$ $\mathtt{execute}$($\square$, $\mathit{id}(c)$, $\mathit{dsc}(c)$, $\mathit{c}(c)$, $\mathit{issuer}(c)$, $\mathit{O}$, $\mathit{t}$, $\mathcal{G}$, $\mathcal{L}$, $i$) = $\bot$,
then the event $e' = \mathtt{Deleted}~\mathit{id}(c)$ is triggered:
$$
\infer[(A), (C), (D'').]{
\langle \mathcal{C} \setminus \{c\}, \mathcal{D}, \mathcal{B}, t, \mathcal{G}, i, \mathcal{L}, e' : \mathtt{E} \rangle
}{
\langle c \in \mathcal{C}, \mathcal{D}, \mathcal{B}, t, \mathcal{G}, i, \mathcal{L}, \mathtt{E} \rangle 
}
$$

\noindent
{\bf [Tick]} The tick rule increments the global time $t$:

$$
\infer[]{
\langle \mathcal{C} , \mathcal{D}, \mathcal{B'}, t + 1, \mathcal{G'}, i', \mathcal{L'}, \mathtt{E} \rangle
}{
\langle \mathcal{C} , \mathcal{D}, \mathcal{B'}, t, \mathcal{G'}, i', \mathcal{L'}, \mathtt{E} \rangle
}
$$

The rules {\bf [Join]}, {\bf [Join {\tt OR}]}, and {\bf [Fail]} cannot be applied in the same time because their side conditions exclude each other. On the other hand, the rules {\bf [Issue]} and {\bf [Tick]} can be applied any time. Also, note that {\bf [Join]}, {\bf [Join {\tt OR}]}, and {\bf [Fail]} are the only rules that actually execute contracts and may produce modifications in the balance, the ledger, and the list of events.

\begin{definition}
The rules above define a relation between states. By $s \step s'$ we denote that there is a step (given by one of the rules {\bf [Issue]}, {\bf [Join]}, {\bf [Join {\tt OR}]}, {\bf [Fail]}, or {\bf [Tick]}) from the state $s$ to the state $s'$. We also use $s \steps s'$ to denote the reflexive and transitive closure of $\step$.
\end{definition}

\subsection{Metaproperties}
\label{sec:metaprops}
\paragraph{State consistency}
We start by defining what conditions need to be satisfied for a state to be consistent.

\begin{definition}
\label{def:consistentstate}
A state $s$ is \emph{consistent} if:
\begin{enumerate}
    \item $\forall c\,.\,c \in \mathcal{C}(s) \to \mathit{fresh}(s) > \mathit{id}(c)$;
    \item $\forall i\,.\, (\mathtt{\small Executed}~i) \in \mathtt{E}(s) \lor (\mathtt{\small Deleted}~i) \in \mathtt{E}(s) \to \mathit{fresh}(s) > i$;
    \item $\forall c\,.\, c \in \mathcal{C}(s) \to ((\mathtt{\small Executed}~\mathit{id}(c)) \not\in \mathtt{E}(s) \land (\mathtt{\small Deleted}~\mathit{id}(c)) \not\in \mathtt{E}(s))$;
    \item $\forall i\,.\, \lnot ((\mathtt{\small Executed}~i)\in\mathtt{E}(s)\land(\mathtt{\small Deleted}~i) \in \mathtt{E}(s)).$
\end{enumerate}
\end{definition}

The first two conditions capture the fact that the id field of the state is always \emph{fresh}, i.e., it always strictly greater than all the other identifiers.
Since we do not use hash functions to compute unique identifiers, we have ensure that our unique identifier generation approach is consistent. 
The last two conditions of Definition~\ref{def:consistentstate} capture the consistency of the generated events. Condition $3$ ensures that an issued contract cannot be executed or deleted, while condition $4$ ensures that a contract cannot be both executed and deleted in the same time.

The first important metaproperty that we prove is that the  $\steps$ relation preserves state consistency:

\begin{theorem}
\label{consistency:state}
For all states $s$, $s'$, such that $s$ is consistent, if $s \steps s'$ then $s'$ is consistent.
\end{theorem}

\noindent
The Coq proof of this theorem is based on several lemmas which correspond to each item in Definition~\ref{def:consistentstate}. These lemmas are proved by induction on $\step$.

\paragraph{Ledger consistency} 
One important property of the ledger is that the registered transactions cannot be removed or modified. In Coq, we are able to prove this for our semantics as well:

\begin{theorem}
\label{consistency:ledger}
For all states $s$ and $s'$, and for all transactions $t$, if $s \steps s'$ and $t \in \mathcal{L}(s)$ then $t \in \mathcal{L}(s')$.
\end{theorem}

\noindent
The proof is based on the fact that the {\tt execute} function only appends transactions to the ledger, and it never modifies the existing transactions in the ledger.

\paragraph{Events consistency}
The $\step$ relation triggers several events: {\tt Executed}, {\tt Deleted}, and {\tt IssuedFor}. 
These events correspond to actions and once an event is triggered it means that some action has been performed. For instance, when a contract is issued an {\tt IssuedFor} event is triggered, and a potential owner can check whether a contract has been issued for him. This action cannot be retracted. In Coq, we prove that triggered events cannot be retracted:

\begin{theorem}
\label{consistency:events}
For all states $s$ and $s'$, and for all events $e$, if $s \steps s'$ and $e \in \mathtt{E}(s)$ then $e \in \mathtt{E}(s')$.
\end{theorem}

\noindent
The proof is based on the fact that the {\tt execute} function only generates new events, and it never modifies the existing list of events.

\paragraph{Other metaproperties}

A very useful property that we intensively use in our Coq proofs is given by the following lemma:

\begin{lemma}
\label{destruct}
For all states $s$ and $s'$, and for all contracts $c$, if $s \step s'$ and $c \in \mathcal{C}(s)$ then $c \in \mathcal{C}(s')$ or $(\mathtt{Executed}~\mathit{id}(c)) \in \mathtt{E}(s')$ or $(\mathtt{Deleted}~~\mathit{id}(c) \in \mathtt{E}(s')$.
\end{lemma}

\noindent
The lemma essentially enumerates the possible outcomes of a step from $s$ to $s'$: either a contract is not executed and it remains available in the contracts of $s'$, or it is executed and a corresponding event was triggered. 
The proof is by induction on $\step$.

Another property that we prove in Coq is that the time cannot go backwards when performing steps:

\begin{lemma}
For all $s$ and $s'$, if $s \steps s'$ then $\mathit{t}(s) \leq \mathit{t}(s')$.
\end{lemma}
\noindent
The proof is by induction on both $\steps$ and $\step$.


\section{Verifying properties of Findel derivatives}
\label{sec:experiments}

In this section we verify the properties \cm, \es, and \as~for several Findel derivatives. 
Sometimes, proofs cannot be completed. In this case we prove a proposition that indicates a vulnerability. 

All the subsequent properties are formalized and certified in Coq. The full Coq codebase is available online at \url{https://github.com/andreiarusoaie/findel-semantics-coq}. 
Since everything is already proved and verified in Coq, we do not show any proofs in this section\footnote{Most of the proofs are by induction on $\step$ or $\steps$, and can be found in the indicated codebase.}.

For convenience, in the rest of this section, we assume that $I$ is always the issuer and $O$ is always the owner. We explicitly mention when their roles change. We also introduce some useful notations and definitions:

\begin{notation}
If {\tt P} is a primitive, then we denote by $c_{\tt P}$ the contract where $\mathit{prim}(c_{\tt P})=\mathtt{P}$, $\mathit{issuer}(c_{\tt P})=I$, $\mathit{owner}(c_{\tt P})=O$, and $\mathit{id}(c_{\tt P})= \mathit{id}(\mathtt{P})$. 
\end{notation}

\begin{definition}
A contract $c$ is \emph{executed between $s_1$ and $s_2$ at time $t$} if $c\in\mathcal{C}(s_1)$, $\mathtt{Executed}~\mathit{id}(c) \in \mathtt{E}(s_2)$, $\mathit{time}(s_1) = t$ and $s_1 \step s_2$.
\end{definition}

\begin{definition}
A contract $c$ is \emph{deleted between $s_1$ and $s_2$ at time $t$} if $c\in\mathcal{C}(s_1)$, $\mathtt{Deleted}~\mathit{id}(c) \in \mathtt{E}(s_2)$, $\mathit{time}(s_1) = t$ and $s_1 \step s_2$.
\end{definition}

\begin{definition}
An owner $O$ \emph{joins a contract $c$ between states $s_1$ and $s_2$ at time $t$} if there are states $s$ and $s'$ such that $s_1 \steps s$, $s' \steps s_2$ and $c$ is executed or deleted between $s$ and $s'$ at time $t$.
\end{definition}


\subsection{Fixed-rate currency exchange}
Recall the fixed-rate currency exchange ($\mathit{FRCE}$) shown in Example~\ref{ex:frce}.
We prove in Coq that $I$ receives 11 dollars from $O$, while $O$ receives 10 euros from $I$:

\begin{proposition}
\label{th:frce2}
For all $s$, $s'$, with $s$ consistent, if O joins $c_\mathit{FRCE}$ between $s$ and $s'$ then there exists $\mathit{tx}_1$ such that \tr{$\mathit{tx}_1$}{I}{O}{10}{USD}{$\mathit{id}(c_\mathit{FRCE})$}$\in \mathcal{L}(s')$.
\end{proposition}

\begin{proposition}
\label{th:frce1}
For all $s$, $s'$, with $s$ consistent, if O joins $c_\mathit{FRCE}$ between $s$ and $s'$ then there exists $\mathit{tx}_2$ such that \tr{$\mathit{tx}_2$}{O}{I}{11}{USD}{$\mathit{id}(c_\mathit{FRCE})$}$\in \mathcal{L}(s')$.
\end{proposition}

In Coq, we actually prove a more general version of these properties, where the amounts are multiplied by the scale of the contract. These propositions guarantee the \cm and \as~properties. \es does not make sense here because no external sources are used. 


\subsection{External rate currency exchange}

A more interesting example is a currency exchange derivative where the exchange rate is provided by an external source, i.e., a gateway:

\noindent
{\small $\mathit{ERCE} \eqbydef$ \tt And (Give (Scale n (One USD)))\\
\hspace*{\fill} (ScaleObs addr (Scale n (One EUR)))}

\noindent
An  interesting question is what happens if the gateway fails to provide an exchange rate $r$? 
First, we prove that the expected transactions are generated if the gateway successfully provides an exchange rate $r$:

\begin{proposition}
\label{th:erce1}
For all $s$, $s'$, with $s$ consistent, if O joins $c_\mathit{ERCE}$ between $s$ and $s'$, and $\mathtt{query}~\mathcal{G}(s)~\mathtt{addr}~t = r$ then there exists $\mathit{tx}_1$ such that \tr{$\mathit{tx}_1$}{I}{O}{$\mathit{sc}(c_\mathit{ERCE}) * r$}{USD}{$\mathit{id}(c_\mathit{ERCE})$}$\in \mathcal{L}(s')$.
\end{proposition}

\begin{proposition}
\label{th:erce2}
For all $s$, $s'$, with $s$ consistent, if O joins $c_\mathit{ERCE}$ between $s$ and $s'$, and $\mathtt{query}~\mathcal{G}(s)~\mathtt{addr}~t = r$ then there exists $\mathit{tx}_2$ such that \tr{$\mathit{tx}_2$}{O}{I}{$\mathit{sc}(c_\mathit{ERCE})$}{EUR}{$\mathit{id}(c_\mathit{ERCE})$}$\in \mathcal{L}(s')$.
\end{proposition}

Second, if the gateway query fails, we prove that the ledger remains unchanged, that is, \es:
\begin{proposition}
\label{th:invalidgtw}
For all $s$, $s'$, with $s$ consistent, if $c_\mathit{ERCE}$ is executed (or deleted) between $s$ and $s'$ and $\mathtt{query}~\mathcal{G}(s)~\mathtt{addr}~t(s) = \mathtt{None}$ then $\mathcal{L}(s) = \mathcal{L}(s')$.
\end{proposition}

\noindent
Thus, no transaction has been performed, and the balance of the users is not affected by the gateway failure.

All the propositions in this section guarantee \cm, \es, and \as for this Findel derivative.


%
%
%
%

\subsection{Zero-coupon bond}

Our next case study is the zero-coupon bond (Example~\ref{ex:zcb}), where the issuer sells a zero-coupon bond that pays 11 {\tt USD} in one year for 10 {\tt USD}:\\[1ex]
\hspace*{-1ex}\begin{minipage}{\textwidth}
\small
\tt 
\begin{tabular}{c c l}
  $ZCB \eqbydef$&\big(And & \\
     && \hspace*{-3ex}(Give (Scale (One USD)  10))\\
     && \hspace*{-3ex}(At (now+t) (Scale (One USD) 11))\big)\\[1ex]
\end{tabular}
\end{minipage}

\noindent
First, we are concerned about the rights of the issuer:

\begin{proposition}
For all $s$, $s'$, with $s$ consistent, if O joins $c_\mathit{ZCB}$ between $s$ and $s'$ then there is a transaction $tx_2$ such that \tr{$tx_2$}{O}{I}{$\mathit{sc}(c_\mathit{ZCB}) * 10$}{USD}{$\mathit{id}(c_\mathit{ZCB})$}$\in \mathcal{L}(s')$.
\end{proposition}

{\tt At} generates a new contract which can be executed between ${\small \tt {now} + t - \delta}$ and ${\small \tt {now} + t + \delta}$ (cf. Table~\ref{tbl:additional}). The owner gets paid only if he joins in this time interval:

\begin{proposition}
For all $s$, $s'$, {\tt now}, {\tt t}, with $s$ consistent, if O joins $c_\mathit{ZCB}$ between $s$ and $s'$, and O joins the contract $c_\mathit{OR}$ generated by $c_\mathit{ZCB}$ at $t' \in [{\small \tt {now} + t - \delta}, {\small \tt {now} + t + \delta}]$ then there is $tx_1$ such that \tr{$tx_1$}{I}{O}{$\mathit{sc}(c_\mathit{OR}) * 11$}{USD}{$\mathit{id}(c_\mathit{OR})$}$\in \mathcal{L}(s')$.
\end{proposition}

\noindent
The hypotheses of the above proposition should be carefully considered:
if $O$ does not join in time, the he does not receive 11 $\mathtt{USD}$.
The next proposition reveals a security vulnerability (time constraints) that affects $O$:

\begin{proposition}
For all $s$, $s'$, {\tt now}, {\tt t}, with $s$ consistent, if O joins $c_\mathit{ZCB}$ between $s$ and $s'$ and O joins the contract $c_\mathit{OR}$ generated by $c_\mathit{ZCB}$ at $t' > {\small \tt {now} + t + \delta}$ then ${\tt Deleted~\mathit{id}(c_\mathit{OR})} \in \mathtt{E}(s')$.
\end{proposition}

\noindent
{\tt Deleted} is generated by the {\bf [Fail]} rule, so the ledger remains unchanged and the owner does not get paid.
So, for $\mathit{ZCB}$ only the properties \cm and \as hold.

\subsection{An option derivative}

Recall the option derivative $\mathit{OPT}$ from Section~\ref{sec:mot}:\\[1ex]
\begin{minipage}{\linewidth}
\small
\tt 
\begin{tabular}{c c l}
  $\mathit{OPT} \eqbydef$&\big(And& \\
     && \hspace*{-2ex}(Before t (Or (Give (One USD)),\\
     && ~~~~~~~~~~~~~(Give (One EUR)))\\
     && \hspace*{-2ex}(After t+2 (Scale 1 (One GBP)))\big)\\[1ex]
\end{tabular}{}
\end{minipage}

We explained in Section~\ref{sec:mot} that Mallory requests and receives 1 {\tt GBP} after {\tt t+2}. This is proved by the next proposition:

\begin{proposition}
For all $s$, $s'$, $t$, with $s$ consistent, if $O$ joins $c_\mathit{OPT}$ between $s$ and $s'$  before $\mathtt{t}$, and then $O$ joins the contract $c_\mathit{After}$ generated by $c_\mathit{OPT}$ at $t'$ where $t' > \mathtt{t + 2}$, then there is $tx_1$ such that \tr{$tx_1$}{I}{O}{$\mathit{sc}(c_\mathit{After})$}{GBP}{$\mathit{id}(c_\mathit{After})$}$\in\mathcal{L}(s')$.
\end{proposition}

This proposition ensures partially (only for the owner) the \cm property. However, it violates \as:

\begin{proposition}
\label{prop:problem}
For all $s$, $s'$, with $s$ consistent before $t$, if $O$ joins $c_\mathit{OPT}$ between $s$ and $s'$ then $O$ is the owner of any contract generated by $c_\mathit{OPT}$.
\end{proposition}

\noindent
So $I$ is not the owner of the generated option derivative, and thus, $O$ is the one who decides whether or not to join the option contract. If $O$ is honest then we can prove:

\begin{proposition}
For all $s$, $s'$, with $s$ consistent, if $O$ joins $c_\mathit{OPT}$ between $s$ and $s'$ before {\tt t}, and then $O$ joins the contract $c_\mathit{Or}$ generated by $c_\mathit{OPT}$ after {\tt t}, then there is $tx_2$ such that \tr{$tx_2$}{O}{I}{$\mathit{sc}(c_\mathit{Or})$}{$\square$}{$\mathit{id}(c_\mathit{Or})$}$\in\mathcal{L}(s')$, where $\square \in \{ \mathtt{EUR, USD} \}$.
\end{proposition}

Indeed, it is not acceptable for $I$ to depend on $O$'s choice.
A possible solution for fixing $\mathit{OPT}$ is to replace {\tt\small (Or (Give (One USD)) (Give (One EUR)))} with {\tt\small (Give (Or (One USD) (One EUR)))}.
Since the {\tt \small Give} primitive swaps the roles of the participants for the enclosed contracts, $I$ becomes the owner, $O$ becomes the issuer, and thus, $I$ can request the payment from $O$. With this change we can prove the next propositions that ensure \cm, \as, and \es for this contract:

\begin{proposition}
\label{prop:problem}
For all $s$, $s'$, with $s$ consistent, if $O$ joins $c_\mathit{OPT'}$ between $s$ and $s'$ before {\tt t}, then $I$ is the owner of the contract $c_\mathit{Or}$ generated by $c_\mathit{OPT'}$ whose primitive is {\tt\small (Or (One USD) (One EUR))} and its issuer is $O$.
\end{proposition}

\begin{proposition}
For all $s$, $s'$, with $s$ consistent, if $O$ joins $c_\mathit{OPT'}$ between $s$ and $s'$ before {\tt t}, and then $I$ joins $c_\mathit{Or}$ generated by $c_\mathit{OPT'}$, then there is $tx_1$ such that \tr{$tx_1$}{O}{I}{$\mathit{sc}(c_\mathit{Or})$}{$\square$}{$\mathit{id}(c_\mathit{Or})$}$\in\mathcal{L}(s')$, where $\square \in \{\mathtt{EUR, USD}\}$.
\end{proposition}


\subsection{Credit Default Swap}
\label{sec:cds}
In this section we prove the properties \cm, \es, and \as for a very popular and real-life financial derivative type: \emph{credit default swap} (CDS). 
This type of derivative enables investors to swap credit risk with another investor. Suppose that Alice buys a financial bond of value {\tt price} from Bob. The maturity of the bond is 3 years. Every year, Bob has to pay Alice a fee {\tt FY}, and {\tt price} when maturity is reached. Alice wants to protect her investment: she  joins a CDS issued by a seller C with a better credit rating than Bob. For this, Alice pays an yearly fee {\tt F} to {\tt C}.  If Bob defaults, then {\tt C} will pay to Alice the {\tt price} and the remaining fees:

\hspace*{-4ex}
\begin{minipage}{\textwidth}
\small
\tt
\begin{tabular}{lcll}
CDS&\hspace*{-3ex}$\eqbydef$&\hspace*{-3ex}(And\\
      &&&\hspace*{-5ex}(* first year *)\\
      &&&\hspace*{-5ex}(And \\
      &&&\hspace*{-2ex}(Give (Scale F (One USD)))\\
          &&&\hspace*{-2ex}(pay\_at\_t (now+1yr) addr\\
          &&&\hspace*{8ex} (price + (2 * FY)))\\
      &&&\hspace*{-5ex})\\
      &&&\hspace*{-5ex}(And\\
          &&&(* second year *)\\
          &&&(yearly\_check (now+1yr)(now+2yrs) \\
          &&&\hspace*{18ex}addr price FY F 1)\\
          &&&(* third year *)\\
          &&&(yearly\_check (now+2yrs)(now+3yrs)\\
          &&&\hspace*{18ex}addr price FY F 0)\\
      &&&\hspace*{-5ex})\\
  &&&\hspace*{-8ex})\\
\end{tabular}
\end{minipage}

~\\
Since Findel does not allow contracts with three parties, we use a gateway available at address {\tt addr} that can tell whether Bob defaulted or not.
Here we take advantage of the compositionality of Findel: {\tt pay\_at\_t} and {\tt yearly\_check} are Findel contracts as well. 
First, {\tt pay\_at\_t} pays {\tt sum} at time {\tt t} if Bob defaulted:\\[-1ex]

\begin{minipage}{.5\textwidth}
\centering
\small
\tt
\begin{tabular}{l}
pay\_at\_t $\eqbydef$ (At t (If addr (Scale sum (One USD)) Zero))\\[1ex]
\end{tabular}
\end{minipage}

\noindent
We prove that {\tt pay\_at\_t} has the following property:

\begin{proposition}
\label{prop:pay}
If Bob defaults at time {\tt t} then the issuer of {\tt pay\_at\_t} pays  {\tt sum} to the owner. Otherwise, no transaction between the involved parties is generated by contract {\tt pay\_at\_t}.
\end{proposition}

Second, {\tt yearly\_check} is more complex:\\[-1ex]

\hspace*{-4ex}
\begin{minipage}{\textwidth}
\small
\tt
\begin{tabular}{lll}
yearly\_check $\eqbydef$At t &\hspace*{-2ex}(If &\hspace*{-2ex}addr Zero\\
         &\hspace*{-2ex}(And &\hspace*{-2ex}(Give (Scale F (One USD)))\\
           &&\hspace*{-2ex}(pay\_at\_t t' addr (price + i * FY))))\\[1ex]
\end{tabular}
\end{minipage}

~\\
The contract describes the obligations that parties have at time {\tt t}: if Bob defaulted at {\tt t},  nothing happens. Otherwise, Alice pays {\tt F} to C, and C pays {\tt price + i * FY} at {\tt t'} if Bob defaults. The contract is intended to be used inside CDS: at {\tt now + 1 year}, Alice pays {\tt F} to C, and at {\tt t' = now + 2 years} C pays {\tt price + 1 * FY} to Alice if Bob defaulted at {\tt t'}. We prove that in Coq:

\begin{proposition}
\label{prop:yearly}
If Bob defaults at time {\tt t}, then {\tt yearly\_check} does not generate any transactions. If Bob does not default at {\tt t} then the issuer receives fee {\tt F} from the owner and a {\tt pay\_at\_at t' ({\tt price + i * FY}) }  contract with the same owner is generated.
\end{proposition}

\noindent
Here, {\tt i} is the number of fees left to be paid until maturity is reached. For instance, if Bob defaults at {\tt now + 1 year} then C pays to Alice {\tt price+2*FY}.

The financial obligations of C to Alice when Bob defaults are summarized by Table~\ref{tbl:cds}. The same table lists the financial obligations of Alice to C. In Coq we prove that the Findel specification of {\tt CDS} ensures that these obligations are fullfilled by the parties. Moreover, we prove that no other finacial obligation is generated by {\tt CDS}. 

The compositional nature of Findel contracts allows us to develop incremental proofs for {\tt CDS}. In our proofs for {\tt CDS} we reuse Propositions~\ref{prop:pay} and~\ref{prop:yearly}. The proofs are broken into smaller pieces, and they become less difficult and easier to manage.

\begin{table}
\centering
\begin{tabular}{|c|c|}
\hline
\bf  Bob defaults at &  \bf Obligations of C to Alice \\
\hline
{\tt now + 1 year} &  {\tt price} + {\tt 2 * FY} \\
{\tt now + 2 years} & {\tt price} + {\tt 1 * FY} \\
{\tt now + 3 years} & {\tt price} \\
\hline
\hline
\bf Timestamp &  \bf Obligations of Alice to C \\
\hline
{\tt now} &  F \\
{\tt now + 1 year} & F \\
{\tt now + 2 years} & F \\
\hline
\end{tabular}
\label{tbl:cds}
\caption{The financial obligations of the parties.}
\end{table}

\section{Conclusions}
\label{sec:conclusion}

The recent developments in the blockchain technologies, especially the support for smart contracts, are a perfect match for financial agreements. 
In particular, it makes sense to have DSLs for financial derivatives that run directly on the blockchain, and thus, they are automatically processed by a decentralized network. These languages are specially designed for people in finance. They can focus on the specification rather than learning how to program smart contracts. 
Expressing financial derivatives in a specialized DSL may be more precise and less error prone than specifying them in a general purpose smart contracts language~\citep{Atzei:2017:SAE:3080353.3080363}. 
Either way, mistakes in contracts can happen. 

Our work is complementary to the efforts in previous research like~\citep{jones2000composing} or~\citep{gai}, where DSLs based on similar primitives as Findel were formalised only with the purpose of making various analyses or estimating derivatives values. Findel derivatives are executed in the blockchain, making derivatives susceptible of several known vulnerabilities of smart contracts.
Our infrastructure is meant to help users to discover such vulnerabilities in their financial derivatives.

\paragraph{Future work}
The main disadvantage of Coq is the fact that it is not fully automatic. Other tools, like K~\citep{k}, can help with automation. However, these are not yet capable to generate certificates. On the other hand it is worth investigating whether these tools are more practical than Coq.

Automation of proofs remains a challenge for our approach. Coq allows users to define the so-called \emph{tactics} which helps with improving the proof language. 
Here we define several tactics that we use to discharge very common proof goals. However, a deeper investigation on how to define specific tactics is required.

Another piece of future work is an investigation on other financial DSLs that run on the blockchain, where proving correctness of the encoded financial agreements could help with finding security vulnerabilities.
\paragraph*{Acknowledgements}
This work was supported by a research grant of the ``Alexandru Ioan Cuza'', University of Ia{\c s}i, within the Research Grants program, Grant UAIC, ctr. no. 6/03.01.2018, code GI-UAIC-2017-08.

\bibliographystyle{unsrtnat}
\bibliography{refs}







\end{document}